# Emergence of coherence in the charge-density wave state of *2H*-NbSe$_2$

U. Chatterjee[1,2*], J. Zhao[2,3], M. Iavarone[4], R. Di Capua[4], J. P. Castellan[1,5], G. Karapetrov[6], C. D. Malliakas[1,7], M. G. Kanatzidis[1,7], H. Claus[1], J. P. C. Ruff [8,9], F. Weber[1,5], J. van Wezel[1,10], J. C. Campuzano[1,3], R. Osborn[1], M. Randeria[11], N. Trivedi[11], M. R. Norman[1] and S. Rosenkranz[1*]

[1]*Materials Science Division, Argonne National Laboratory, Argonne, IL 60439, USA.*

[2]*Department of Physics, University of Virginia, Charlottesville, VA 22904, USA.*

[3]*Department of Physics, University of Illinois at Chicago, Chicago, IL 60607, USA.*

[4]*Department of Physics, Temple University, Philadelphia, PA 19122, USA.*

[5]*Institute of Solid State Physics, Karlsruhe Institute of Technology, P.O. Box 3640, D-76021 Karlsruhe, Germany.*

[6]*Department of Physics, Drexel University, Philadelphia, PA 19104, USA.*

[7]*Department of Chemistry, Northwestern University, Evanston, IL 60208, USA.*

[8]*Advanced Photon Source, Argonne National Laboratory, Argonne, IL 60439, USA.*

[9]*CHESS, Cornell University, Ithaca, NY 14853, USA.*

[10]*Institute for Theoretical Physics, University of Amsterdam, Tyndall Avenue, 1090 GL Amsterdam, the Netherlands.*

[11]*Department of Physics, Ohio State University, Columbus, OH 43210, USA.*

**A charge-density wave (CDW) state has a broken symmetry described by a complex order parameter with an amplitude and a phase. The conventional view, based on clean, weak-coupling systems, is that a finite amplitude and long-range phase coherence set in simultaneously at the CDW transition temperature $T_{cdw}$. Here we investigate, using photoemission, X-ray scattering and scanning tunneling microscopy, the canonical CDW compound *2H*-NbSe$_2$ intercalated with Mn and Co, and show that the conventional view is untenable. We find that, either at high temperature or at large intercalation, CDW order becomes short-ranged with a well-defined amplitude that impacts the electronic dispersion, giving rise to an energy gap. The phase transition at $T_{cdw}$ marks the onset of long-range order with global phase coherence, leading to sharp electronic excitations. Our observations emphasize the importance of phase fluctuations in strongly coupled CDW systems and provide insights into the significance of phase incoherence in 'pseudogap' states.**

Correspondence and requests for materials should be addressed to U.C. (uc5j@virginia.edu) or S.R. (srosenkranz@anl.gov)

The formation of charge-density waves (CDWs) and superconductivity are archetypical examples of symmetry breaking in materials, which are characterized by a complex order parameter. In clean weak-coupling systems, the formation of the amplitude and the establishment of macroscopic phase coherence are known to occur simultaneously at the transition temperature[1,2], but the situation may be dramatically different at strong coupling or in the presence of disorder[3-9]. Such systems generally exhibit short correlation lengths and a transition temperature that is greatly suppressed from the expected mean-field value. This opens the possibility for a gap in the electronic spectra to persist in the absence of long-range order over a large temperature range and the opportunity to study this so-called pseudogap behaviour, which has been observed in a wide range of systems from high-temperature superconductors[10-12] to disordered superconducting thin films[13] and cold atoms[14], in a simple system.

In the CDW state below the transition temperature $T_{cdw}$, a modulation appears in the density of conduction electrons that is accompanied by a periodic lattice distortion[1]. Although this lattice distortion leads to an increase in the elastic energy, there is a net gain in the free energy of the system due to a reduction in the energy of the electrons via the opening of an energy gap $\Delta$. The CDW state is characterized by the order parameter $\delta\rho(\mathbf{r}) = \rho(\mathbf{r}) - \rho_{av}(\mathbf{r}) = \rho_0 \cos[\mathbf{q}\cdot\mathbf{r} + \phi(\mathbf{r})]$, where $\mathbf{q}$ is the CDW wave vector, $\rho_{av}$ the average charge density, $\rho_0$ the amplitude of the CDW order parameter, which is proportional to the energy gap $\Delta$, and $\phi$ the phase of the CDW, i.e., the location of the charge modulation with respect to the underlying lattice. It is straightforward to realize that there are two ways to destroy the order parameter — (i) by reducing the amplitude $\rho_0$ through excitations across the energy gap and/or (ii) by randomization of the phase $\phi$ either through thermal or quantum fluctuations. Similar to the case of BCS transitions in clean superconductors, the expectation value of $\delta\rho$ in weakly coupled CDW systems vanishes via (i), as, in general, (ii) is not relevant due to the large magnitude of the phase stiffness energy[1,3]. However, strongly coupled CDW systems can be susceptible to strong phase fluctuations due to their shorter coherence lengths[1,4-7].

To explore this systematically as a function of disorder in a material with a simple electronic and crystal structure, we investigate pristine and intercalated (with Co and Mn ions) $2H$-NbSe$_2$ samples, where CDW ordering involves strong electron-phonon coupling[15,16]. In our investigations, we employ a combination of experimental probes. Scanning tunneling microscopy (STM) and X-ray diffraction (XRD) are used to measure the real and momentum space structure of the CDW order, respectively. Angle resolved photoemission spectroscopy (ARPES) is employed to investigate the presence of an energy gap in the electronic spectra, which is proportional to the amplitude of the order parameter, and to infer the existence or disappearance of phase coherence of the order parameter through the presence or absence of sharp coherence peaks in the ARPES spectra. Furthermore, $T_{cdw}$ is determined by tracking the onset of the CDW order parameter in XRD and the CDW-induced anomaly in transport measurements. Our main results from this extensive set of measurements are as follows. First, the temperature $T_{cdw}$, above which CDW phase coherence is destroyed, is suppressed with intercalation $x$ and vanishes at a quantum phase transition (QPT) at $x_c$. Second, the CDW state below $T_{cdw}(x)$ and $x < x_c$ has an energy gap and sharp electronic excitations. Finally, the CDW energy gap survives above $T_{cdw}(x)$ and $x > x_c$ in the state which has only short-range CDW correlations, but the electronic excitations are no longer well-defined.

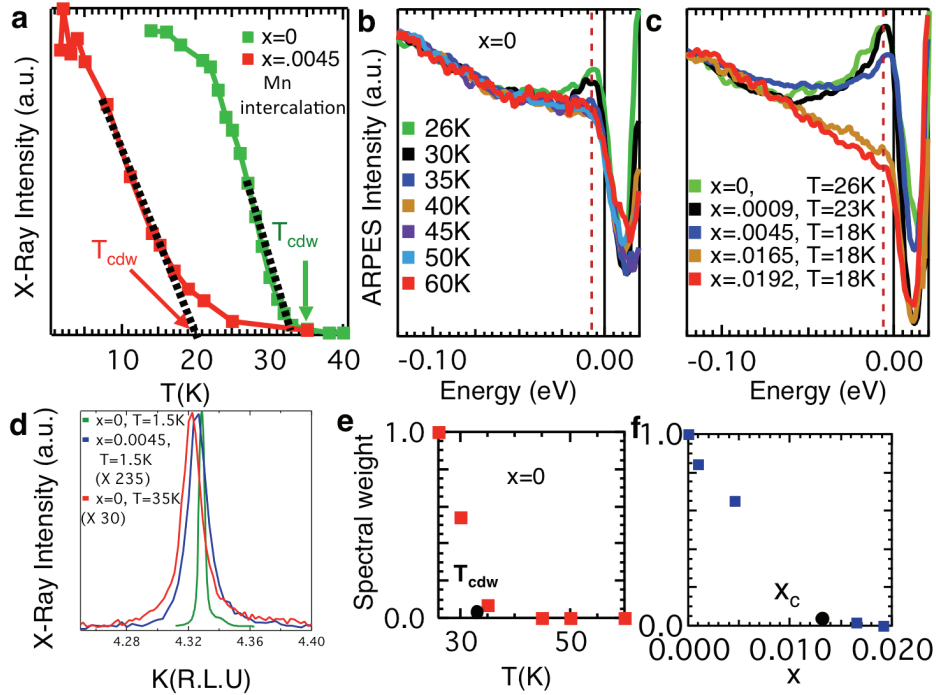

**Figure 1 | CDW phase transition for pure and intercalated 2H-NbSe$_2$ samples.** (**a**) Intensity of the CDW superstructure peak as a function of $T$ for an $x$=0 (green) and $x$=0.0045 (Mn-intercalated, red). Dotted lines are guides to the eye indicating the growth of the CDW order parameter. Arrows mark the estimated $T_{cdw}$, which is the interpolation of the linear part to zero. (**b**) FF divided EDCs for various temperatures at the momentum location shown by the red dot in Fig. 2d. (**c**) FF divided EDCs for different $x$ at the lowest measured temperatures. The sample with $x$=0.0165 is intercalated with Co, the others are intercalated with Mn. The corresponding momentum location is shown by the pink dot in Fig. 2d. The red dashed line in **b** and **c** shows the energy location of the peak (for $T$< $T_{cdw}$ & $x$<$x_c$) and kink (for $T$>$T_{cdw}$ or $x$>$x_c$) in the EDCs, while the black line is the chemical potential. (**d**) Intensity profile of CDW superstructure peaks along (0,$K$,0) for $x$=0, 0.0045 (intercalated with Mn) at 1.5 K and for $x$=0 at $T$ = 35 K. The intensities have been normalized to that of the (0,4,4) Bragg peak and have subsequently been multiplied by the factors displayed in the legend to make them all visible on the same scale. (**e**,**f**) Normalized spectral weight associated with the coherence peak, following the procedure as described below, as a function of $T$ for $x$=0 and as a function of $x$ at the lowest measured temperature. The black dots correspond to $T_{cdw}$ for $x$=0 and $x_c$ ~ 0.013, respectively. For $x$=0, the coherent spectral weight was obtained as the remaining integrated spectral intensity after subtracting the highest measured temperature (60K) as background and dividing by the value obtained at the lowest measured temperature. We adopt the same procedure, as for the normalized spectral weight as a function of $T$, for its $x$ dependence by considering the spectrum at the highest value of $x$ as the background and dividing the coherent spectral weight at the lowest measured temperature for each $x$ by the one for $x$=0.

### Results

**XRD and transport.** Previous investigations of the CDW transition in $2H$-NbSe$_2$ using neutron and XRD have shown that a superstructure with incommensurate wave vector $q=(1-\delta)a^*/3$, where $a^*=4\pi/\sqrt{3}a$, lattice parameter $a=3.44$ Å and $\delta\sim 0.02$, appears below $T_{cdw}=33$K (ref. 17). However, the origin of the CDW and the exact nature of the Nb atom motion involved were only recently established using inelastic X-ray scattering[16] and superspace crystallography[18], respectively. To explore the evolution of the CDW transition with increasing disorder, we use synchrotron XRD and transport measurements on a number of intercalated samples (for details, see Methods section and Supplementary Fig. 1). Figure 1a shows the intensity of the incommensurate superstructure peaks, which is proportional to the square of the CDW order parameter. For the undoped compound, the superstructure intensity increases continuously from the temperature-independent background present above $T_{cdw}$, as expected for a second-order phase transition. Moreover, the width of this superstructure peak, which is a measure of the inverse of the CDW correlation length, becomes resolution limited below $T_{cdw}$, implying the presence of long-range order (Fig. 1d). For intercalated samples, the XRD superstructure peak exhibits a reasonably sharp onset of intensity (Fig. 1a) as a function of temperature, but its width (Fig. 1d) is not resolution limited even at our lowest measured temperature. This suggests that the process of intercalation gives rise to disorder that impacts the CDW state, consistent with previous work on $2H$-NbSe$_2$ and other dichalcogenide systems[19-22]. We can nevertheless still identify a CDW transition temperature, even in the intercalated, disordered samples, from a linear extrapolation of the temperature dependence of the superstructure intensity, as indicated in Fig. 1a.

From our XRD data on samples with different concentrations $x$ of intercalant ions, we observe that $T_{cdw}$ is quickly suppressed (Fig. 1a) as $x$ is increased and beyond $x_c$, superstructure peaks become very broad, indicating that the CDW order becomes short-range. Qualitatively, similar attributes have been identified in resistivity measurements as well: the CDW induced anomaly in the resistivity becomes weaker with increasing $x$, shifts to lower temperatures and eventually disappears beyond $x_c$ (see ref. 22, Supplementary Note 1 and Supplementary Fig. 2). We note that the effect of doping on the CDW is the same irrespective of whether the intercalating ion is Mn or Co as observed by STM, discussed later. The disappearance of $T_{cdw}$ naturally points to the existence of a QPT at $x_c$. Although our measurements show clear signatures of the change in the CDW order across $x_c$, as expected at a QPT, we have not been able to characterize this QPT in greater detail due to the difficulty in synthesizing samples with dense-enough $x$ values around $x_c$.

**Angle-resolved photoemission.** To elucidate the evolution of the electronic structure across $T_{cdw}$ and $x_c$, we now focus on the $T$- and $x$-dependent ARPES measurements. We have taken data over a large region of momentum space in the first Brillouin zone (Supplementary Note 2 and Supplementary Fig. 3a). However, we will predominantly be concerned with ARPES data as a function of energy at specific momentum values close to the high symmetry $K$-$M$ line (Fig. 2d). The reason for choosing this particular direction is that the CDW gap is maximum along it and hence easily detectable by ARPES[23,24]. The Fermi function (FF) at corresponding temperatures is divided out to better visualize the existence of the gap $\Delta$ from such energy distribution curves (EDCs)[25,26]. In Fig. 1b, we display the spectral function obtained from the FF-divided EDCs as a function of $T$ for $x$=0 (raw data, shown in the Supplementary Fig. 3b,c, agree with previous ARPES studies[23,24]). At the lowest temperature ($T$=26K < $T_{cdw}$=33K), the spectrum exhibits a coherence peak at an energy below the chemical potential $\mu$, signalling a non-zero value of the gap. Unlike superconducting or one-dimensional CDW systems, we find that this gap is centred at positive energies, i.e., above $\mu$. This particle-hole asymmetry is clearly evident from the FF-divided ARPES data shown in Fig. 1b, which has a negative slope at $\mu$. Consequently, the estimation of the energy gap based on the leading edge or by symmetrizing the spectra around $\mu$[23,24] underestimates the size of the gap. Given that the minimum of the FF-divided ARPES intensity occurs above $\mu$, we can not determine the exact magnitude of the energy gap, but can only establish its existence and that, in agreement with recent STM data[27], it is larger than estimated by symmetrisation and leading-edge analysis in previous ARPES studies. The presence of a particle-hole asymmetric gap is an indication that Fermi surface nesting is not important for driving the CDW in 2$H$-NbSe$_2$. This is supported by recent measurements of soft phonon modes using inelastic X-ray scattering[16]. However, differing conclusions about the role of nesting have been offered in previous ARPES studies[28-30].

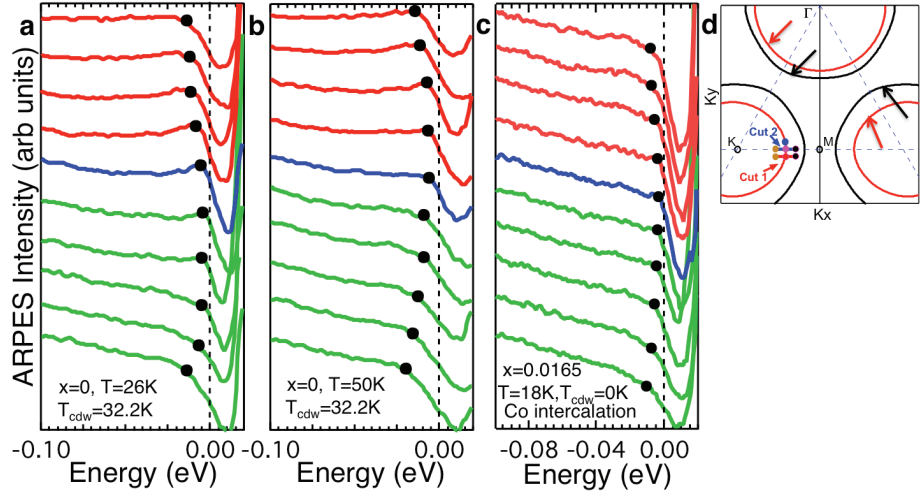

**Figure 2 | Electronic dispersion exhibiting back bending.** (**a,b**) FF divided EDCs as a function of momentum **k** along cut 1 (defined in panel **d**), which is close to the $K$-$M$ line, for $x=0$ and at $T=26$K and $T=50$K, respectively. Red EDCs are associated with momenta **k** inside the inner Fermi surface (centred around the $K$ point) starting from the brown dot (shown in **d**) to the red dot along cut 1, green ones to the remaining **k** points along cut 1 between the red and black dots. (**c**) FF divided EDCs for $x=0.0165$ (Co intercalation) at $T=18$K, along cut 2, which is along $K$-$M$, as shown in **d**. Red EDCs correspond to **k** starting from the brown dot inside the inner Fermi surface (centred around the $K$ point) to the pink dot along cut 2, green ones to the remaining **k** points between the pink and black dots. Blue curves indicate EDCs at the Fermi momentum, the black dots show the locations of the peak and kinks of the EDCs, and the black dashed lines are at the chemical potential. The method for locating the black dots is described in Supplementary Note 5 and Supplementary Fig. 7. (**d**) Fermi surface plot based on the tight-binding fit to the band structure of $2H$-NbSe$_2$ in ref. 24. Black and red arrows denote outer and inner Fermi surface barrels.

With increase in temperature, the intensity of the coherence peak is reduced, but its energy remains the same. Finally, above $T_{cdw}$, this peak disappears and remarkably, the spectra become almost $T$ independent. Although the spectra for $T > T_{cdw}$ do not have a discernable peak, they do have a well-defined "kink", i.e., a sharp change of slope. Closer inspection of Fig. 1b reveals that the location of this kink in the spectra above $T_{cdw}$ is similar to that of the coherence peaks below $T_{cdw}$, as shown by the red dotted line. Although we cannot obtain the exact magnitude of the energy gap as a function of $T$ through $T_{cdw}$, the fact that peak/kink structure of the ARPES spectra lies below the chemical potential clearly demonstrates that the energy gap persists for $T > T_{cdw}$. On the other hand, there is a loss of phase coherence as indicated by the absence of a peak in these spectra. Our ARPES data on 2$H$-TaS$_2$ exhibits similar features (Supplementary Note 3 and Supplementary Fig. 5), which in turn suggests that these features are generic to quasi two-dimensional CDW systems. We have also seen similar results for other values of $x$ as well in 2$H$-NbSe$_2$ (Supplementary Fig. 4). Therefore, $T_{cdw}$ can be identified as the transition from an incoherent, gapped electronic state to a coherent one. This implies that although the CDW order parameter vanishes above $T_{cdw}$, its amplitude does not.

In Fig. 1e, we display the normalized spectral weight as a function of $T$ (the details of the normalization procedure is described in the caption of Fig. 1). One can clearly observe that it decays monotonically with $T$ and vanishes above $T_{cdw}$. The $x$ dependence of the spectra (Fig. 1c) through $x_c$ resembles the $T$ dependence through $T_{cdw}$, e.g., the spectrum at the lowest measured temperature progressively loses its peak with increasing $x$ and eventually becomes featureless for $x > x_c$. It can readily be seen (Fig. 1f) that the $x$ dependence of the normalized coherent spectral weight follows a similar trend as the $T$ dependence, that is, it steadily goes down with increasing $x$ and vanishes above $x_c$. However, a finite energy gap persists for all $x$, even beyond $x_c$. In other words, there exists an extended region in the $(x,T)$ phase plane of 2$H$-NbSe$_2$ systems where the spectral function has an energy gap but no coherence.

We now look at the details of the $(x,T)$ phase plane in the context of pairing. As a direct consequence of particle-hole coupling, the electronic dispersion, i.e., the relation between electronic energy and momentum, is altered from the one in the normal state. In the CDW state with ordering wave vector $\mathbf{q}$, an electronic state with momentum $\mathbf{k}$ gets coupled to another with momentum $\mathbf{k+q}$, and the dispersion is modified to $E_\mathbf{k} = \frac{1}{2}(\varepsilon_\mathbf{k}+\varepsilon_\mathbf{k+q}) \pm [\frac{1}{4}(\varepsilon_\mathbf{k}-\varepsilon_\mathbf{k+q})^2 + \Delta_\mathbf{k}^2]^{1/2}$, where $\Delta_\mathbf{k}$ is the energy gap and $\varepsilon_\mathbf{k}$ the normal state dispersion[1,6]. The dispersion of the lower branch deviates from $\varepsilon_\mathbf{k}$ as $\varepsilon_\mathbf{k}$ approaches $\varepsilon_\mathbf{k+q}$, reaches a maximum when $\varepsilon_\mathbf{k}=\varepsilon_\mathbf{k+q}$, and then bends back to follow $\varepsilon_\mathbf{k+q}$ instead (Supplementary Fig. 6 and Supplementary Note 4). This bending back effect, commonly known as the Bogoliubov dispersion[2], occurs in the superconducting state due to electron-electron pairing and has been directly observed by ARPES in cuprates[31-33]. As shown in Fig. 2a, by monitoring the peak positions and kinks of the EDCs (see Supplementary Note 5 and Supplementary Fig. 7 for details), it is easy to see that the electronic dispersion below $T_{cdw}$ for $x$=0 has the expected bending back behaviour. And quite remarkably, the bending back of the dispersion is visible even when the CDW is short-ranged—(i) in the $x$=0 sample at $T\sim$50K > $T_{cdw}$ in Fig. 2b and (ii) an intercalated sample with $x\sim$0.0165 > $x_c$ at $T$=18K in Fig. 2c. The presence of CDW gap below and above $T_{cdw}$ along with the persistence of a back bending feature in the electronic dispersion naturally suggests that the electron-hole pairs induced by the CDW persist above $T_{cdw}$ and beyond $x_c$. We point out that similar observations have been made in the pseudogap phase of cuprates where, just as in the superconducting state, a Bogoliubov dispersion was found to exist above $T_c$, which was considered as evidence for pairing without phase coherence[34]. Taking into consideration of all these observations, one can realize that for $T>T_{cdw}$ or $x > x_c$ in the $(x,T)$ phase plane of 2$H$-NbSe$_2$, there are electron-hole pairs without coherence, which in turn implies that CDW order parameter vanishes because of phase incoherence.

**Scanning tunnelling microscopy.** So far, we have concentrated on the structure and the relevant electronic excitations associated with the CDW order in momentum space. To visualize them in real space, we have performed high-resolution STM measurements as a function of $T$ and $x$ as shown in Fig. 3, where the upper panels are STM topographic images, each of which covers an area of 18.5 nm by 18.5 nm, the second row shows Fourier filtered images, and the bottom panels are 2D-Fourier transform (FT) images. Figure 3a corresponds to the $x$=0 sample at $T$=4.2 K. The almost perfect hexagonal lattice pattern confirms that our pristine samples are of very high quality and practically defect free. A close inspection of the topographic image clearly reveals the presence of a locally commensurate CDW pattern that repeats after every third lattice spacing; however, phase slips lead to the slight incommensuration observed in XRD as well as in the FT of the STM image shown in Fig. 3i[35]. The long-range order of both the lattice as well as the CDW is reflected in the sharp peaks observed in the FT image, where the outermost set of the six strong peaks correspond to the Bragg peaks of the underlying lattice, and the innermost ones at ~ 1/3 the wave vector correspond to the lowest order CDW superlattice peaks (Fig. 3, Fig. 4). Our topographic as well as FT images of pure samples are consistent with earlier STM work[27,36-38]. While the CDW order is well visible in the topographic image of the pure sample, the CDW corrugation in NbSe$_2$ is very small and to enhance the subtle features due to CDW modulations, we show Fourier-filtered images in the middle row, panels **e-h** of Fig. 3. These images are obtained by applying a mask to the FT image that only retains the region around the lowest-order CDW peaks, e.g. the region between the dash-dotted circles shown in Fig. 3i, and transforming back to real space. This results in images of the CDW maxima and allows a direct visualization of the phase and CDW domain structure. To ensure that the filtered images are free from possible artifacts due to the filtering procedure, we compared images obtained using different sizes of the filtering mask. The main features of the filtered images remain the same regardless of the filter used. The long-range order and phase coherence of the CDW in the pure sample is clearly reflected in its Fourier-filtered image, where an almost perfect CDW lattice is obtained (Fig. 3e). STM topographic images of intercalated samples reveal the presence of defects, both at the surface and beneath the surface, as bright spots. In the sample with nominal doping $x$=0.0045 (Fig. 3b), the local density of defects in the topographic image is in the range $0.006<x<0.0075$. In contrast to the pure sample, the Fourier filtered image (Fig. 3f) reveals that coherent CDW order does not exist uniformly over the entire sample, but only in the form of patches. As a demonstration, we have indicated one such region with locally well-defined CDW order in the Fourier-filtered image (Fig. 3f) by a dashed circle. In other words, CDW order in the intercalated sample with $x$=0.0045 lacks-long range translational order, which in turn supports the observation, as in Fig. 1, that the CDW order in this intercalated sample is not long range. Consistent with this, the CDW peaks of the FT image in Fig. 3j are broader than those of the pure sample (Figs 3i and 4). As $x$ is increased to 0.04, the size of the patches with coherent CDW order (Figs. 3d,h) is further reduced and the peaks in the FT become even broader, leading to a ring-like structure (Fig. 3l). In Fig. 4a, we present the line profiles of the FT in Fig. 3i-l along the atomic wave vectors as a function of doping, and in Fig. 4b we display the Gaussian fits to these CDW peaks (Supplementary Note 6), from which we clearly see the broadening of the CDW peak with doping. These results simply relate to the fact that the CDW coherence length decays with increasing $x$ and eventually becomes short-range, consistent with our ARPES and XRD observations. We note that these observations are independent of the intercalating ion, see Supplementary Note 7 and Supplementary Figs 8,9. STM observations of domain-like structures having well defined CDW order, similar to those shown in Fig. 3f, have been reported in a number of disordered dichalcogenide systems[39,40].

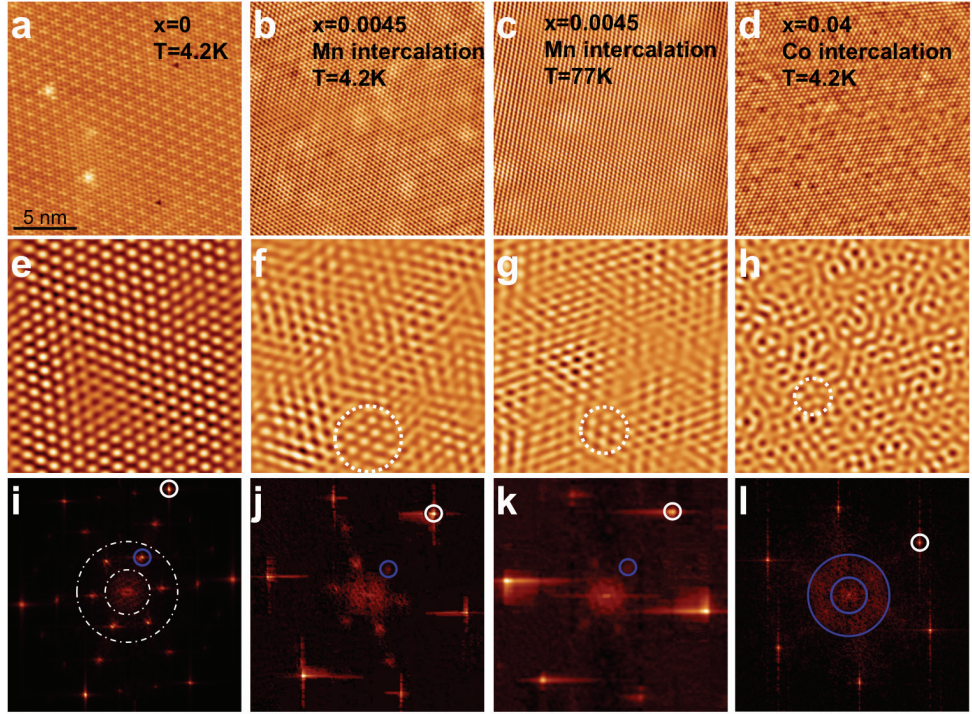

**Figure 3 | STM data on pristine and intercalated 2H-NbSe$_2$ samples.** STM topographic images of (**a**) $x$=0 at 4.2 K, (**b,c**) $x$=0.0045 (Mn intercalated) at 4.2 K and 77 K, and (**d**) $x$=0.04 (Co intercalated) at 4.2 K. The scan areas are 18.5 nm x 18.5 nm. The images have been acquired at V = +50 mV and $I$ = 100 pA. The total z-deflection is 0.6 Å in **a,c**, 1.0 Å in **b**, and 1.2 Å in **d**. (**e-h**) Fourier filtered images obtained by retaining only the region around the lowest-order CDW peak in the FT images, for example, the region $0.2 q_{BRAGG} < q < 0.45 q_{BRAGG}$ indicated by the white dash-dotted circles in **i**, and transforming back to real space. The dashed white circles in these filtered images show examples of patches with well-defined CDW structures in the intercalated samples. (**i-l**) 2D-FTs of the topographic images in **a-d**. The white circles indicate the position of one of the lowest-order Bragg peaks at $q_{BRAGG}$, blue circles in **i-k** at ~ 1/3 $q_{Bragg}$ indicate the position of the lowest order CDW superlattice peaks. In samples with high CDW disorder, the FT does not show clear CDW peaks but rather a ring-like structure, as indicated by the blue circles in **l** for Co$_{0.04}$NbSe$_2$.

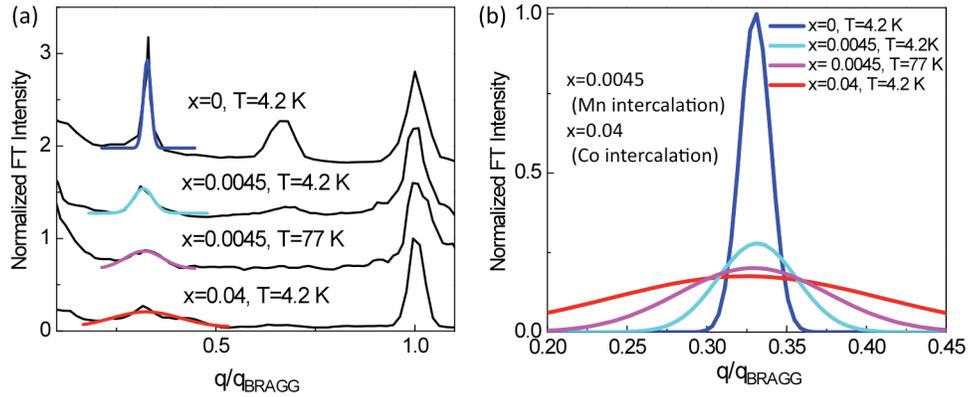

**Figure 4 | STM FT Profiles.** (**a**) Line profiles of the FT images in Fig. 3i-3l along one of the lattice wavevector directions. The curves have been shifted for clarity. The intensity of each profile has been normalized to the intensity of the Bragg peak. The CDW peak at $q=q_{BRAGG}/3$ has been fitted with a Gaussian and the fit is superimposed on the FT line profile for each curve. (**b**) Gaussian fits of the CDW peaks in **a** normalized to the CDW peak of pure NbSe$_2$.

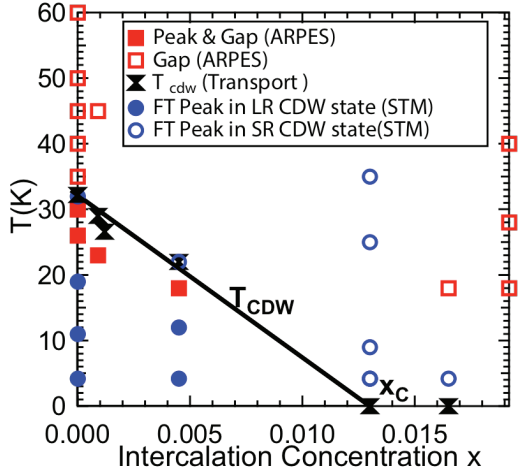

**Figure 5 | Intercalation versus temperature diagram of CDW correlations in 2H-NbSe$_2$.** Overview plot of the results from ARPES, STM, transport, and XRD measurements. ARPES data points are shown by red squares. Filled squares correspond to the simultaneous presence of the coherence peak and an energy gap, empty ones to the presence of only an energy gap. STM data points are shown by blue circles, indicating the presence of a CDW. Bow ties denote $T_{cdw}$, obtained from transport measurements, marking the transition from a long-range ordered CDW state to a short-range ordered state as a function of $x$ and $T$. Extrapolation of this phase line leads to $x_c \sim 0.013$, beyond which there is only short-range order at all measured temperatures. The superconducting transitions, observed for all samples except the very highest doping as listed in Supplementary Table 1, are not shown in this diagram.

We now focus on STM data for $T > T_{cdw}$ (Fig. 3c,g,k) corresponding to the $x=0.0045$ sample at $T=77$K $> T_{cdw}=20$K. Although weak in intensity, well-defined CDW modulations survive in various regions, as best seen in the Fourier-filtered image Fig. 3g, and the FT image (Fig. 3k) still clearly shows broad CDW peaks, indicating the persistence of short-range CDW correlations to temperatures well above $T_{cdw}$. In the pure sample, we observe very weak CDW modulations localized around the impurities up to 45 K (Supplementary Figure 10a). Given the small amount of defects and the weak modulation, only very weak CDW peaks are observed in the 2D-FT. We do not observe any sign of CDW modulations in real space or in 2D-FT at 77 K (Supplementary Fig. 10b). This can be understood by realizing that the STM images represent a time-averaged view of the sample, and that in regions far from any pinning defects the signatures of short-range order are averaged out by dynamic fluctuations.

## Discussion

The phase diagram of the CDW correlations in intercalated 2H-NbSe$_2$ shown in Fig. 5 summarizes our results (for simplicity, the superconducting transition observed in all but the very highest doped sample are not shown in this figure). The squares are based on the ARPES spectra that are: (1) coherent and gapped (solid squares) within the triangular region where the CDW order is present, (2) incoherent and gapped (empty squares) outside of this triangle, indicating short-range CDW order instead. The circles correspond to FT-STM data —solid ones within the triangular region where long range CDW order exist and empty ones outside where CDW order is short range. The bow ties indicate the CDW transition temperature determined from transport measurements, and are included for completeness.

This phase diagram closely resembles that of the transition from the superconducting to the pseudogap state in cuprates[41], and that of the superconducting to insulating transition in disordered systems[9,13,42]. The experimental observation of the persistence of an energy gap and the disappearance of single-particle coherence across classical or quantum phase transitions seems to be a common feature of these entirely different systems. The enigmatic pseudogap phase of cuprates is in many ways the most complicated of these examples, with close proximity to a Mott insulator and various competing orders. In this respect, the simplicity of 2H-NbSe$_2$ studied in this paper makes it possible to focus on a single order parameter, associated with CDW formation, and examine in detail how the spectroscopic signatures evolve across the QPT where this ordering is destroyed.

## Methods

**Intercalated 2H-NbSe$_2$ samples.** Cobalt (Co) and manganese (Mn) have been intercalated in pristine 2H-NbSe$_2$ single crystals by adopting the procedures explained in refs 43,44. XRD patterns show that our intercalated samples have the same structure as the pristine 2H-NbSe$_2$ samples. The concentration of Co and Mn were determined from energy-dispersive X-ray microanalysis measurements. While performing energy-dispersive X-ray microprobe analysis on intercalated samples, we have randomly chosen different regions of the samples and found that the atomic concentrations of Co and Mn remain almost the same, meaning there is no evidence for clustering on the length scale of 100 nm. This is substantiated by previous EXAFS measurements[45].

The superconducting critical temperature ($T_c$) for pristine 2H-NbSe$_2$ samples is ~7.2K, and it was previously shown that $T_c$ gets suppressed due to intercalation[46]. Our magnetization versus temperature measurements for different samples show sharp superconducting transitions (Supplementary Figure 1), which further support that our intercalated samples are homogeneous. Supplementary Table 1 lists details of all the samples used in this study.

**Transport measurements.** The temperature-dependent resistivity of the single crystals was measured using regular four-terminal transport measurement technique performed with a Physical Property Measurement System by Quantum Design equipped with external device control option. Keithley 6220 current source and Keithley 2182 nanovoltmeter were used to apply a DC current and measure the voltage drop, respectively.

**XRD measurements.** Synchrotron XRD experiments were performed at the Advanced Photon Source, Argonne National Laboratory. Data were taken in transmission geometry with 22.3keV X-rays on the 11-ID-D beamline using a Ketek Silicon Drift point detector and on the high-energy station of beamline 6-ID-D with 80keV X-rays and a Pilatus 100K area detector. In both setups, the samples were mounted on the cold head of a closed-cycle helium displex refrigerator and sealed inside a Be-can with He exchange gas.

**ARPES measurements.** ARPES measurements were carried out at the PGM beamline of the Synchrotron Radiation Center, Stoughton, WI, and at the SIS beamline of the Swiss Light Source, Paul Scherrer Institut, Switzerland, utilizing Scienta R4000 analyzers and 22 eV photons for all the measurements. Typical spot sizes of the synchrotron beam on the samples surface is ~50-150 microns times 300 microns.

**STM measurements.** Low temperature STM measurements have been performed using a Unisoku UHV STM system (Unisoku USM-1300 combined with RHK electronics), with a base pressure of $1.0 \times 10^{-10}$ Torr.

The samples were cleaved just before cooling down. Atomically resolved images were acquired in the constant current mode with a constant voltage between sample and tip. We used Pt-Ir tips in all of our experiments.

**Acknowledgements.**
Work at Argonne (U.C., J.Z., J.P.C., C.D.M., M.G.K., H.C., J.P.C.R., F.W., J.C.C, R.O., M.R.N., S.R.) was supported by the Materials Science and Engineering Division, Basic Energy Sciences, Office of Science, U.S. Dept. of Energy. Work at Temple University (M.I.) and Drexel University (G.K.) was supported as part of the Center for the Computational Design of Functional Layered Materials, an Energy Frontier Research Center funded by the U.S. DOE, BES under Award DE-SC0012575. J.v.W. acknowledges support from a VIDI grant financed by the Netherlands Organization for Scientific Research (NOW). M.R. was supported by the DOE-BES grant DE-SC0005035. NT was supported by the U.S. DOE, Office of Science, Grant DE-FG02-07ER46423. The Synchrotron Radiation Center is supported by the University of Wisconsin, Madison. Synchrotron x-ray scattering experiments were carried out at the Advanced Photon Source, which is supported by the DOE, Office of Science, BES. We thank Ming Shi for his support with the experiments at the Swiss Light Source, Paul Scherrer Institut, Switzerland, and D. Robinson and K. Attenkofer for their support with the X-ray diffraction measurements at the Advanced Photon Source, Argonne National Laboratory.

**Author Contributions.**
U.C. and S.R. proposed the research project, U.C. and J.Z. carried out the ARPES measurements and analysis of the ARPES data. J.P.C., J.P.C.R., F.W., R.O. and S.R. carried out the XRD measurements and analysis of the XRD data. M.I., R.D.C., and G.K. carried out the STM, transport measurements and analysis of STM and transport data. M.I., G.K., C.D.M. and M.G.K. synthesised high quality pristine and intercalated NbSe$_2$ samples. H.C. conducted the superconducting SQUID measurements on the samples to determine their superconducting critical temperatures. U.C., J.v.W., M.R., N.T., M.R.N., and S.R. wrote the manuscript. All authors discussed the results and contributed to the manuscript.

**Additional Information**
**Supplementary Information** attached

**Competing financial interests:** The authors declare no competing financial interest

**Supplementary Information**

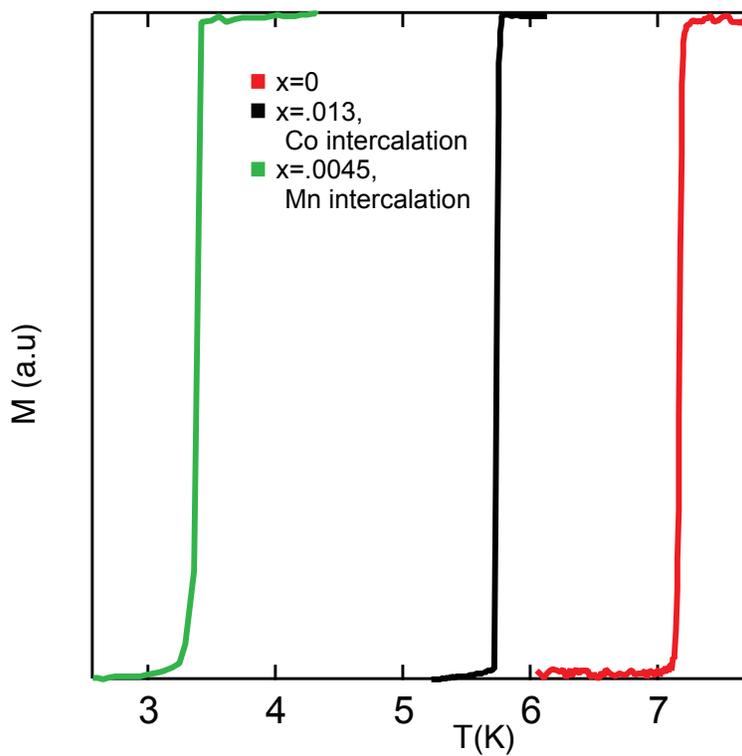

**Supplementary Figure 1 | Magnetization versus temperature curves for pristine and intercalated *2H*-NbSe$_2$ samples.** Superconducting transition for different values of *x*.

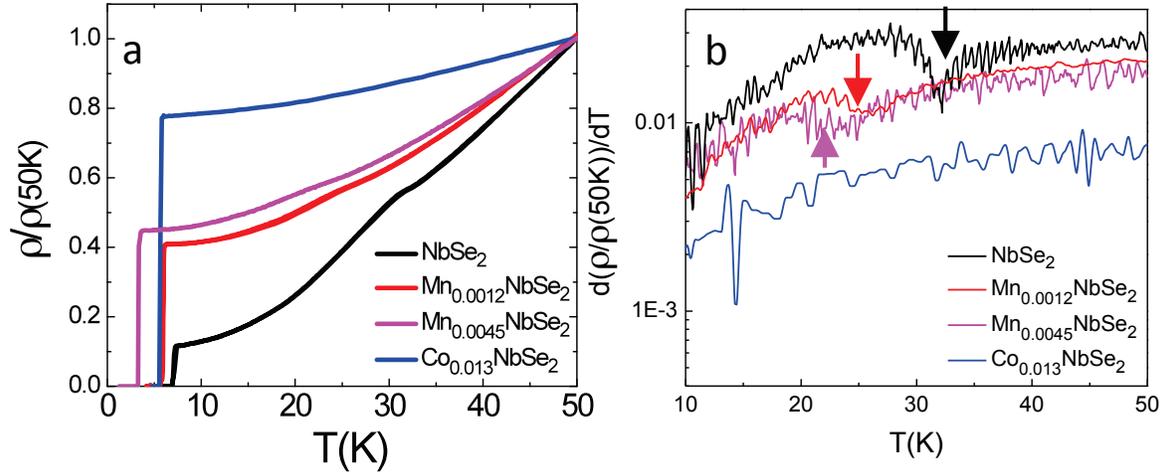

**Supplementary Figure 2 | Transport data on pure and intercalated samples.** (a) Resistivity normalized to its value at 50K for $x$=0, 0.0012 (Mn intercalated), 0.0045 (Mn intercalated) and 0.013 (Co intercalated). (b) Temperature derivative of the $\rho$ vs $T$ plot of the data presented in (a). Arrows indicate the CDW transition temperature. The sample with $x$=0.013 does not show any appreciable change in the first derivative of the resistivity data.

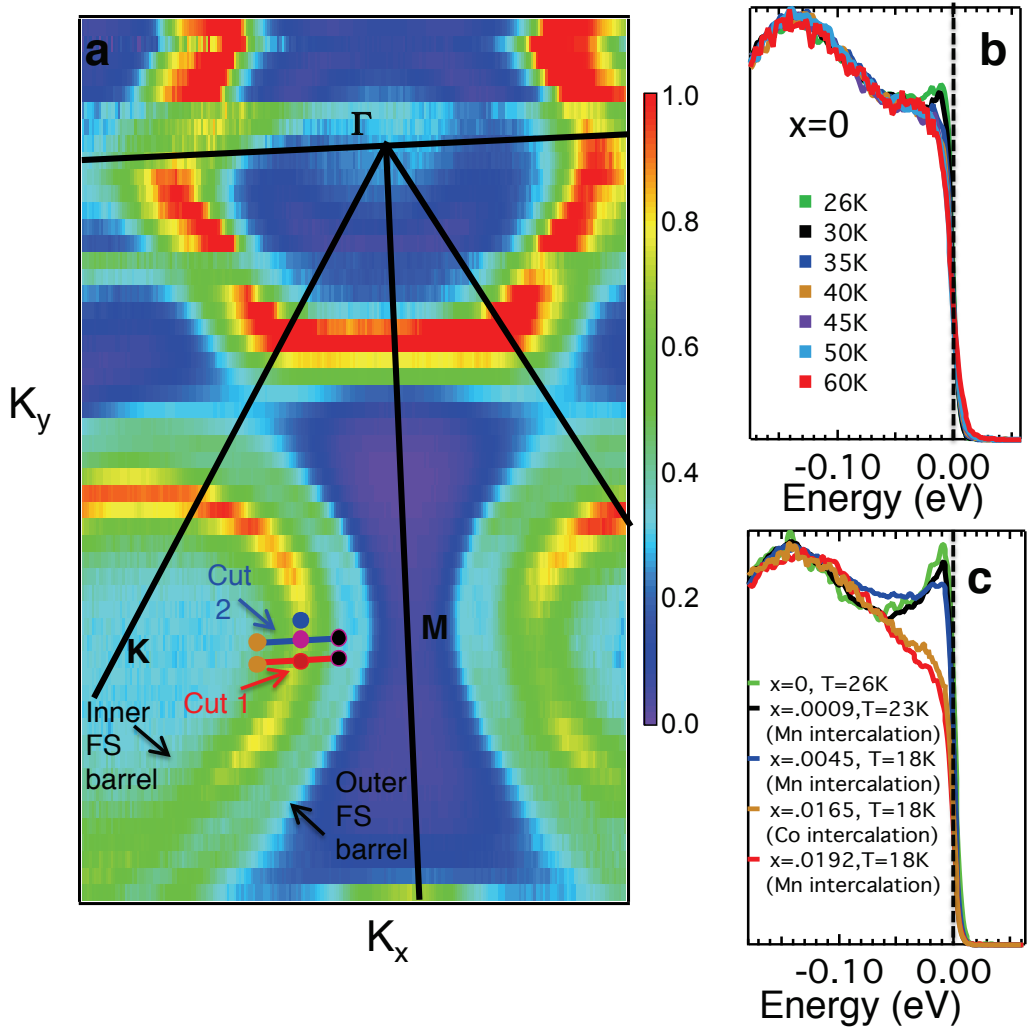

**Supplementary Figure 3 | Raw ARPES data:** (a) Raw ARPES intensity map at zero binding energy (integrated over an energy window of 10 meV) as a function of $k_x$ and $k_y$ for an $x$=0 sample exhibiting double walled Fermi surfaces centered around the $\Gamma$ as well as the K point. As shown in Fig. 2d, cut 1 is along the red line, while cut 2 is along the blue line. Data in Figs. 2a, 2b are taken along cut 1 and Fig. 2c along cut 2. Raw EDCs corresponding to the Fermi function divided ones in (b) Fig. 1b and (c) Fig. 1c. Black dotted lines denote the location of the chemical potential.

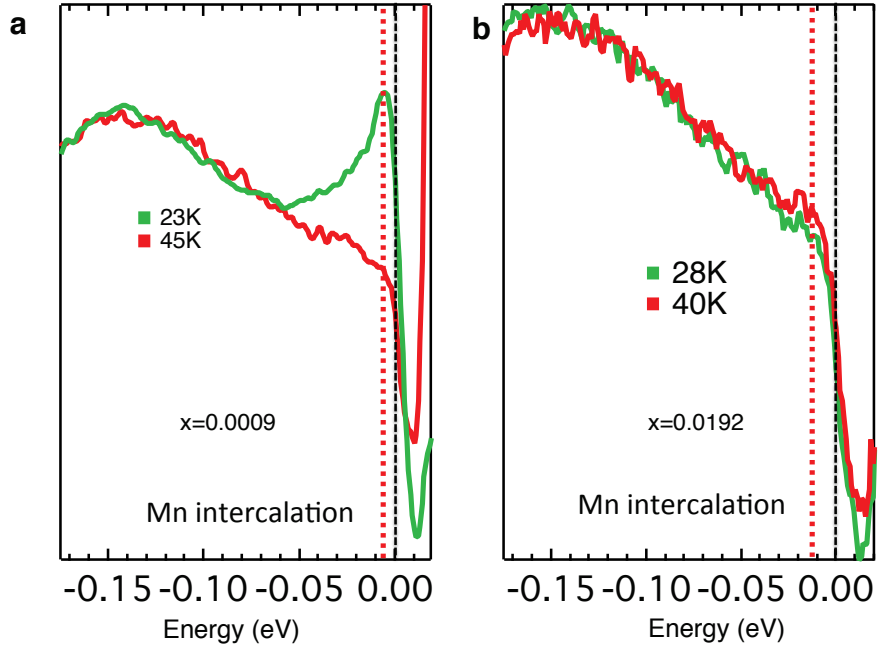

**Supplementary Figure 4 | Fermi function divided ARPES data for two different intercalated samples.** Temperature dependent Fermi function divided ARPES EDCs for (a) $x$=0.0009 (taken at the red dot in Fig. 2d) and (b) $x$=0.0192 (taken at the blue dot in Fig. 2d). Black dotted lines denote the location of the chemical potential. Red dotted lines are used to indicate the energy gap defined as the location of either the peak or kink (i.e., discontinuous change in slope) in the spectrum.

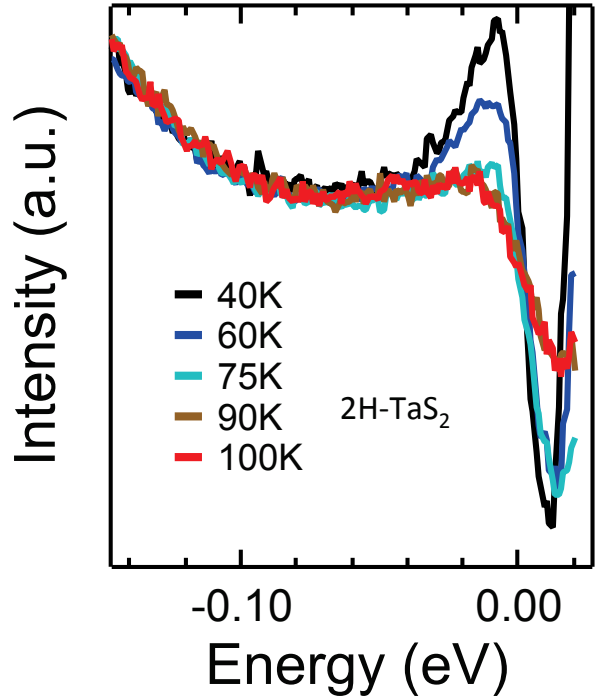

**Supplementary Figure 5 | Fermi function divided ARPES data from *2H*-TaS$_2$.** Temperature dependent Fermi function divided ARPES EDCs for *2H*-TaS$_2$. ARPES data correspond to the momentum location at which the K-M line intersects the Fermi surface. Like in *2H*-NbSe$_2$ (Fig. 1b of main manuscript), here one can also identify a gap and peak in the spectrum for $T<T_{cdw}$ ($T_{cdw}\sim70K$) but only a gap for $T>T_{cdw}$.

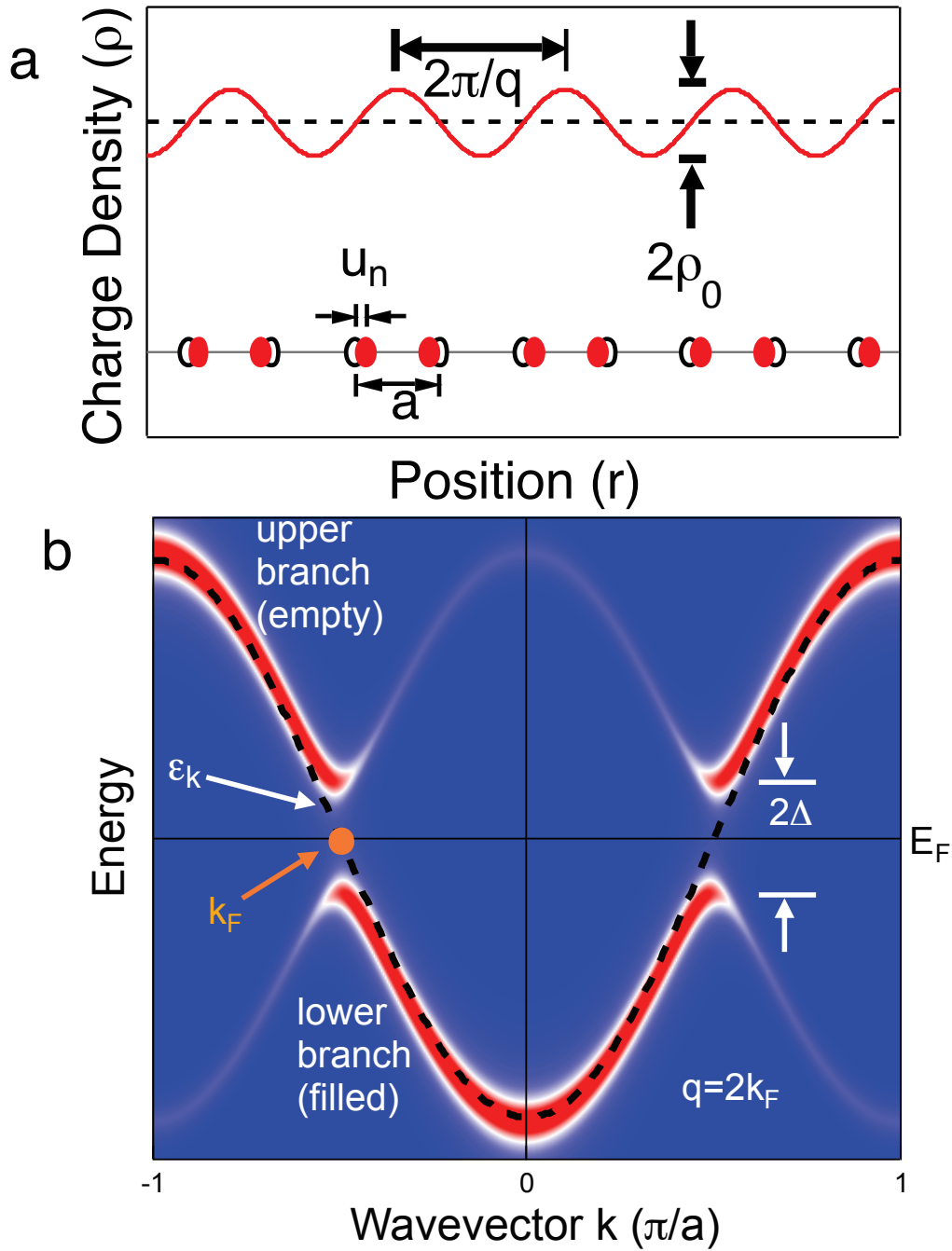

**Supplementary Figure 6 | Electronic dispersion in the CDW state associated with a Peierls distortion in a one-dimensional system.** (a) A cartoon of one-dimensional CDW order in real space and (b) the corresponding electronic dispersion in momentum space in the normal state (black dotted line) and in the CDW state (red lines).

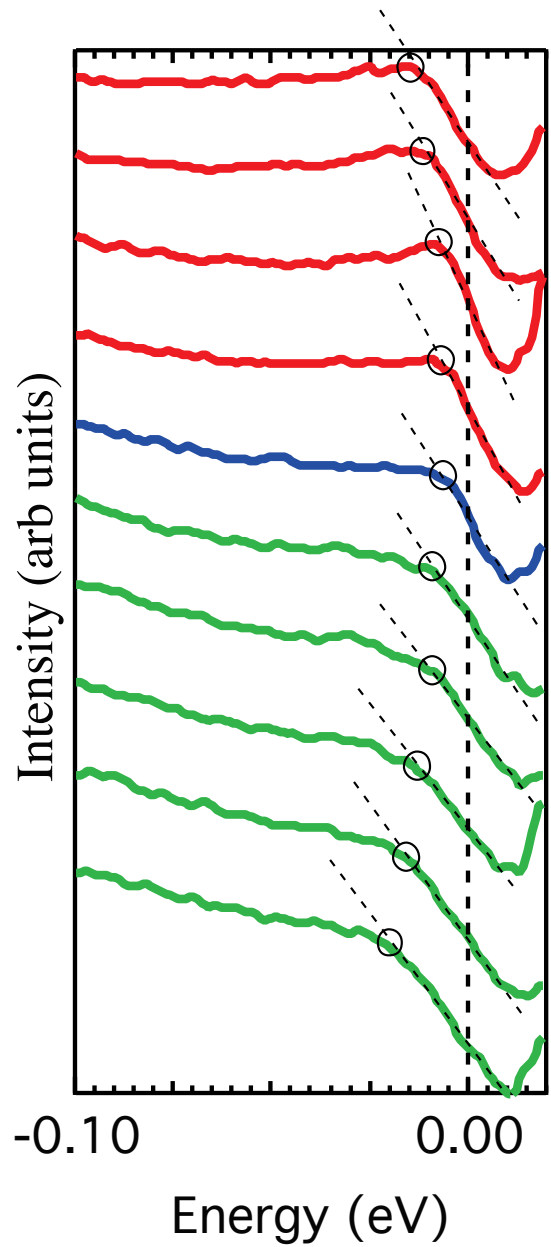

**Supplementary Figure 7 | Identification of peaks and kinks in the ARPES spectra.** Demonstration of how dots are selected, particularly when the spectra are not associated with sharp peaks, to investigate bending back of the electronic dispersion in Fig. 2. In particular, we have chosen Fig. 2b in which peak structures of the spectra are not very clear.

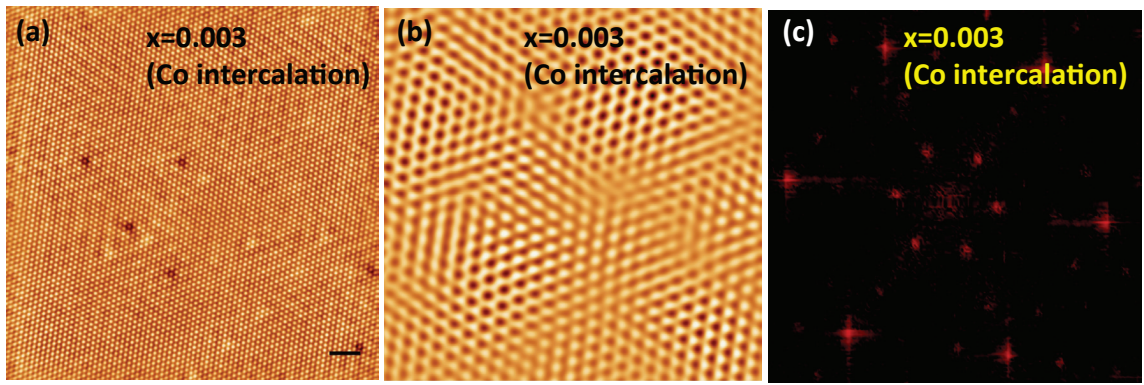

**Supplementary Figure 8 | STM of Co$_{0.003}$NbSe$_2$.** (a) STM topography image acquired with $V$=50 mV and $I$=100 pA. Scan area is 24 nm x 24 nm (scale bar: 2nm). (b) Fourier filtered image of the topography shown in (a). (c) 2D-FT of the image in (a).

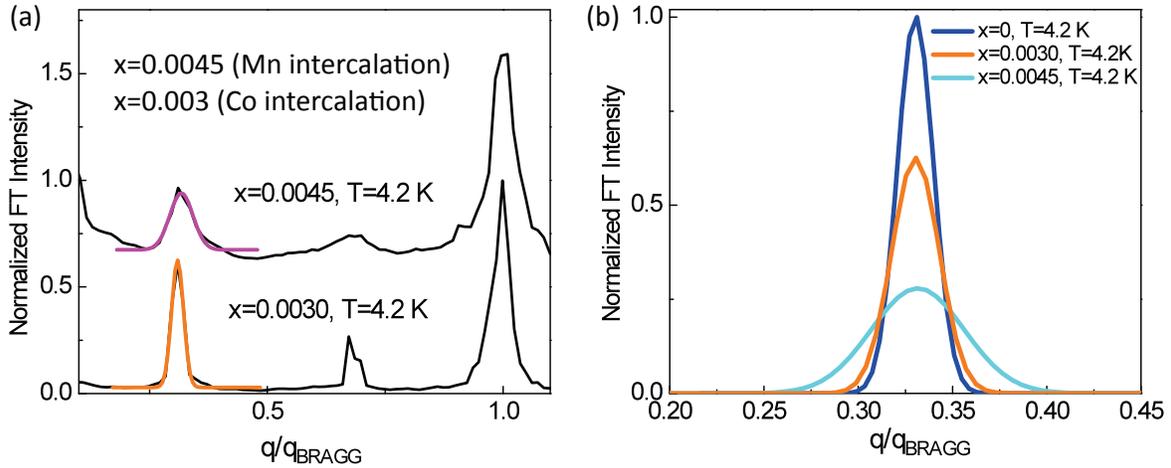

**Supplementary Figure 9 | STM Fourier Transform Profiles for different dopants.**
(a) Line profiles of the FT of the $Mn_{0.0045}NbSe_2$ image (Fig. 3(j) of main text) and the FT of the $Co_{0.003}NbSe_2$ image (Supplementary Figure 8c), along one of the lattice wavevector directions. The curves have been shifted for clarity. The intensity of each profile has been normalized to the intensity of the Bragg peak. The CDW peak at $q=q_{BRAGG}/3$ has been fitted with a Gaussian and the fit is superimposed on the FT line profile for each curve. (b) Gaussian fits of the CDW peaks in $Mn_{0.0045}NbSe_2$ and $Co_{0.003}NbSe_2$ compared to the CDW peak in pure $NbSe_2$.

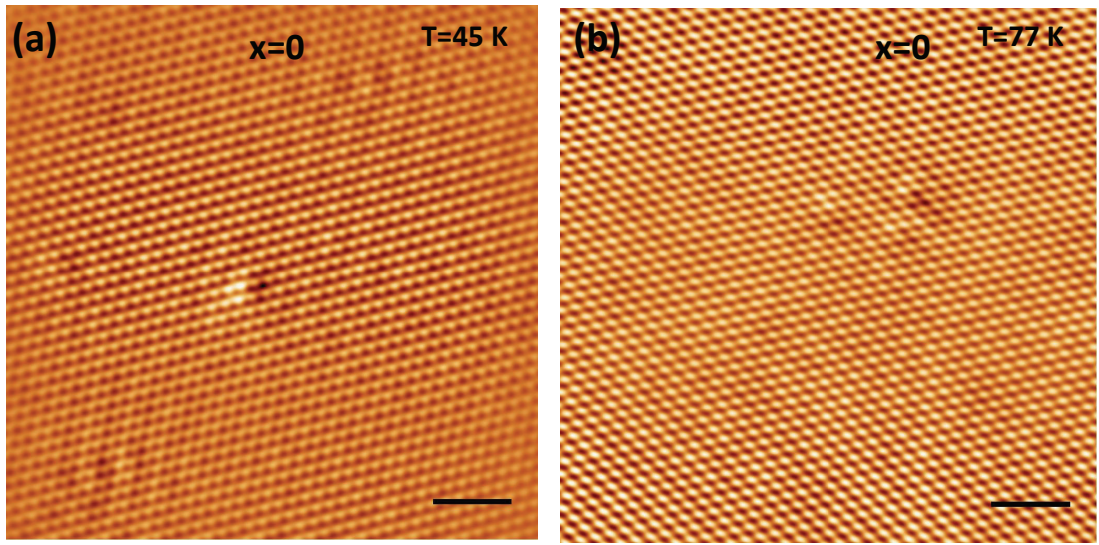

**Supplementary Figure 10 | STM topography of *2H*-NbSe$_2$.** STM topography images of pure NbSe$_2$ above the temperature for long range order. Scan areas for both images are 13.5 nm x 13.5 nm (scale bars: 2 nm). Tunneling conditions are: $V$=50 mV and $I$ =100 pA. (a) $T$=45 K, (b) $T$=77 K.

| $x$ | Type of intercalating ion | $T_c$ | $T_{cdw}$ |
| --- | --- | --- | --- |
| 0 | None | 7.2K | 32.2K |
| 0.0009 | Mn | 6.5K | 29K |
| 0.0012 | Mn | 5.6K | 26.6K |
| 0.0045 | Mn | 3.4K | 23K |
| 0.0192 | Mn | <1.8K | 0K |
| 0.013 | Co | 5.6K | 0K |
| 0.0165 | Co | 4.7K | 0K |
| 0.04 | Co | 3.1K | 0K |

**Supplementary Table 1 | Sample details**

Concentration of the intercalation ion ($x$), type of the intercalating ion, superconducting critical temperature ($T_c$) and $T_{cdw}$ for the samples used in this paper.

**Supplementary Note 1**

**Determination of $T_c$**

We determine the CDW transition temperature $T_{cdw}$ from the CDW induced anomaly in the transport measurements and from the temperature dependence of the CDW order parameter obtained from XRD measurements. We find that these different measurements provide values for $T_{cdw}$ that are close to each other. Supplementary Figure 2 shows resistivity $\rho$ vs temperature $T$ for four samples with $x=0$; $x=0.0012$, $x=0.0045$ (intercalated with Mn); and $x=0.013$ (intercalated with Co), normalized to their individual resistivity at 50K. The derivatives $d\rho/dT$ vs $T$ (Supplementrary Figure 2b) for the three samples with $x < x_c$ exhibit clear anomalies, which we identify as the CDW transition temperature $T_{cdw}$. These transition temperatures agree with the transition temperatures determined as described in the main text from the temperature dependence of the superlattice reflections measured with XRD. The fourth sample with $x > x_c$ does not show any anomaly.

**Supplementary Note 2**

**Raw ARPES data**

Supplementary Figure 3a shows the raw ARPES intensity map at zero binding energy (integrated over an energy window of 10 meV) as a function of $k_x$ and $k_y$. Raw ARPES spectra corresponding to the Fermi function divided ones in Figs. 1b,c are displayed in Supplementary Figures. 3b,c. One can readily observe the energy bands that form two barrels around the K point – the low energy peak/kink corresponds to the inner barrel while the high binding energy "hump" corresponds to the outer barrel (Fig. 2d, Supplementary Figure 3a). For our discussions, we have focused on data in the vicinity of the inner barrel. Supplementary Figure 3b shows a quasiparticle peak that appears at positive binding energy, gets weaker with $T$, and eventually disappears above $T_{cdw}$. Similarly, the spectra in Supplementary Figure 3c lose their peak with increasing $x$, and beyond $x_c$ the peak vanish.

Figure 1b showed the temperature evolution of ARPES spectra across the CDW phase transition. Supplementary Figure 4a shows similar data, i.e. Fermi function divided EDCs taken at the red dot in Figs. 2d, for $x=0.0009$ (Mn intercalation) at temperatures below and above $T_{cdw}$. The low temperature spectrum has both an energy gap and a peak, while for $T >$ $T_{cdw}$, the peak disappears but the energy gap remains. Supplementary Figure 4b shows Fermi function divided EDCs at $T$=28K and 40K for $x$=0.0192 (Mn intercalation) > $x_c$, taken at **k** marked by the blue dot in Fig. 2d. In contrast to Fig. 1b and Supplementary Figure 4a, there is no signature of a peak in Supplementary Figure 4b even at the lowest measured temperatures. Rather there is a kink (marked by the red dotted line) at positive binding energy, with the minimum of the spectrum above the chemical potential, again indicating the presence of an energy gap.

**Supplementary Note 3**

**Loss of coherence at the CDW phase transition in 2*H*-TaS$_2$**

2*H*-TaS$_2$, like 2*H*-NbSe$_2$, exhibits CDW ordering, but with higher transition temperature ($T_{cdw}$ ~70K) (Ref. 1). Our ARPES data measured on 2*H*-TaS$_2$ (Supplementary Figure 5) shows similar features as that for 2*H*-NbSe$_2$. The data for $T<T_{cdw}$ consist of a coherence peak as well as an energy gap, while only an energy gap exists for $T>T_{cdw}$. Using TEM measurements, it has been shown that CDW fluctuations persist for temperatures much higher than $T_{cdw}$ (Ref. 2). All this suggests that the pseudogap phase induced by phase incoherence is generic to quasi two-dimensional CDW systems.

**Supplementary Note 4**

**Energy gap and backbending of the electronic dispersion in the CDW state**

In contrast to superconductors where there is electron-electron pairing, a CDW state is associated with a coupling between electrons and holes, and this is manifested in the electronic dispersion, i.e. the relation between the electronic energy and momentum, below $T_{cdw}$. As a consequence, the electronic dispersion in the CDW state gets modified from the one in the normal state, i.e. for $T>T_{cdw}$, and this can be written as[3,4]:

$$E_{\mathbf{k}} = \tfrac{1}{2}(\varepsilon_{\mathbf{k}}+\varepsilon_{\mathbf{k}+\mathbf{q}}) \pm [\tfrac{1}{4}(\varepsilon_{\mathbf{k}}-\varepsilon_{\mathbf{k}+\mathbf{q}})^2 + \Delta_{\mathbf{k}}^2]^{1/2}$$

where $\Delta_{\mathbf{k}}$ is the momentum dependent energy gap, $\varepsilon_{\mathbf{k}}$ is the normal state dispersion ($T>T_{cdw}$) and **q** is the CDW wavevector. For simplicity, we consider CDW ordering in a one-dimensional system, noting that there are some important differences between CDW order in 1D and 2D systems, like the fact that the energy gap in the 2D case may be centered away from the Fermi level (as we observe to be the case in 2*H*-NbSe$_2$). Our primary focus here is to illustrate the salient experimental signatures associated with the formation of charge order irrespective of dimensionality. The most important of these are the opening of a gap and the related backbending of the electronic dispersion, accompanied by coherence factors, which indicate the amount of mixing between the coupled bands that influence the intensity of the experimental signal. These features are manifested in the 1D case in precisely the same manner as they are in a 2D charge ordered state.

Supplementary Figure 6a shows a schematic of the real space structure of a 1D CDW state, which can be understood as a Peierls instability with a CDW wave vector $\mathbf{q}=2\mathbf{k}_f$, where $\mathbf{k}_f$ is the Fermi momentum. The electronic dispersion of this CDW state, using the equation above, is displayed as the red curves in Supplementary Figure 6b. $E_\mathbf{k}$ consists of two branches, which instead of crossing the chemical potential bend downwards/upwards at $\mathbf{k}_f$. This characteristic bending back of the dispersion provides a direct signature of the electron-hole coupling in the CDW state, with the difference in energy between the lower and the upper branches at $\mathbf{k}_f$ being twice the energy gap, $\Delta$.

**Supplementary Note 5**

**Identification of peaks and kinks in ARPES data**

The black dots in Fig.2 correspond to either peaks or the locations of discontinuity, i.e., kinks, in the Fermi function divided ARPES spectra. The procedure for determining the location of the black dots in Figure 2 of the main manuscript is as follows: (i) Whenever there is a discernable peak in the spectrum, it is quite straightforward to select the location of the peak and the black dot corresponds to the peak position. (ii) When there is no well-defined peak we use a simple method: we approximate the leading edge of the spectrum by a straight line and the dot corresponds to that particular energy value at which the straight line starts to deviate from the spectrum. We have considered the data displayed in Fig. 2b, in which the peaks of the spectra are not pronounced, and demonstrated how to determine the locations of black dots in Supplementary Figure 7.

**Supplementary Note 6**

**STM Fourier Transform profiles**

In Fig. 3 the 2D-FT of the STM topographic images for samples with different intercalation density are shown. All the images show hexagonal spots corresponding to the Bragg peaks and the CDW peaks. In Figure 4a, line cuts of the FT along the lattice ordering vector are reported. The intensity of each line cut has been normalized to the intensity of the Bragg peak. In order to quantify the broadening of the CDW peak with the density of intercalating ions (Mn/Co) and the temperature, we fitted the CDW peak with a Gaussian function of the form: $y = y_0 + \left(\dfrac{A}{w\sqrt{\pi/2}}\right)\exp(-2x^2/w^2)$, where $y_0$ is the background of the FT line cut and $w$ is a width parameter. Figure 4b shows the CDW peak Gaussian fits for the different line cuts. From this figure we observe that the intensity of the CDW peak decreases and the peak broadens with increasing intercalant doping and with increasing temperature. The broadening of this peak measures the loss of translational order.

In the case of pure *2H*-NbSe$_2$ above the temperature for long range order, we do observe weak CDW modulations localized around impurities (Supplementary Figure 10a). However, this modulation is very weak and the peaks in the FT are not clearly defined. At *T*=77 K we do not observe any CDW modulations in real space and any CDW peaks in the FT within our resolution, given the low defect density of our samples (Supplementary Figure 10b).

**Supplementary Note 7**

**Effect of different dopants on the CDW order of intercalated *2H*-NbSe$_2$**

Co and Mn intercalated *2H*-NbSe$_2$ crystals with similar doping levels have been characterized with STM. In Supplementary Figure 8 an STM topography image is reported for Co$_{0.0030}$NbSe$_2$, together with the FT and the Fourier filtered image. The Fourier filtered image reveals that the CDW lattice forms patches very similar to those obtained in Mn$_{0.0045}$NbSe$_2$. The line profile of the FT in Supplementary Figure 8c along a crystal lattice direction is compared with the one obtained for Mn$_{0.0045}$NbSe$_2$ (Supplementary Figure 9a), and the fits of the CDW peaks for the two samples are reported in Supplementary Figure 9b. These measurements show that the effect of Co and Mn on the CDW order is the same.